\documentclass{article}
\newcommand{\be}{\begin{equation}}
\newcommand{\ee}{\end{equation}}

\begin{document}

\begin{center}\LARGE \textbf{Unified Description of Plausible Cause and Effect of the Big
Bang} \\ \normalsize \vspace{2em} \Large{D.C. Choudhury} \\
 \normalsize{Department of Physics, Polytechnic University,
Brooklyn, New York 11201 USA
\\ e-mail: dchoudhu@duke.poly.edu}
\end{center}

\begin{quote}
Plausible cause and effect of the big bang model are presented
without violating conventional laws of physics. The initial
cosmological singularity is resolved by introducing the
uncertainty principle of quantum theory. We postulate that,
preceding the big bang at the end of the gravitational collapse,
the total observed mass including all forms of energy of the
universe condensed into the primordial black hole (PBH) in a state
of isotropic and minimal chaos (i.e., nearest to the absolute zero
temperature). The frozen energy of the collapsed state is
quantized such that each quantum of frozen energy (the cosmic
particle) is characterized by the fundamental constants of the
general theory of relativity and of quantum theory. The minimum
size and minimum lifetime of the cosmic particle are estimated
within the framework of the uncertainty principle. It is
considered that these cosmic particles are bosons with intrinsic
spin zero. The minimum size and thermodynamic properties of the
PBH are estimated within the framework of the Bose-Einstein
statistics by taking into account the total observed mass of the
universe. The mechanism of the big bang is given and the currently
accepted Planck temperature at the beginning of the big bang is
predicted. The critical mass density, the mean mass density, and
the missing mass are calculated as a function of the Hubble time
of expansion. Our calculations show that the universe is closed
but expanding. However, the universe remains inside the event
horizon of the primordial black hole.

Subj- class: cosmology: theory - pre- big bang physics; post- big
bang effect.
\end{quote}

\vspace{2em}

\noindent {\textbf{1. INTRODUCTION}} \vspace{1ex}

\noindent \normalsize Our current understanding of the evolution
of the universe is primarily based on the standard big bang model,
an explosion at Planck temperature $\sim 10^{32}$~K (Planck 1899).
The important observational evidences in support of this model
are:  (i) The universe is undergoing an isotropic homogeneous
expansion, a fact first established by Hubble (1929). This has
been experimentally verified in almost all the studies of
thousands of galaxies across the universe with minimal
uncertainty; (ii) Since the early 1940's nuclear physics began to
play a significant role in cosmology when it was realized (Gamow
1946, 1948; Alpher, Bethe, \& Gamow 1948; Alpher, Follin, \&
Herman 1953) that if our universe began with extremely high
densities and temperatures in excess of $10^{12}$~K,
nucleosynthesis would have occurred. The calculated primordial
abundances of the light elements D, $^3$He, $^4$He, and $^7$Li, in
the early universe, within the framework of nuclear reactions and
the standard model of particle physics, are in strikingly good
agreement with the observed data (Boesgaard \& Steigman 1985);
(iii) The presence of the cosmic microwave background radiation
(CMBR), first discovered by Penzias \& Wilson (1965); followed by
the most spectacular confirmation of its existence which came from
the observation of the 2.72-K CMBR of NASA's Cosmic Background
Explorer (COBE) Satellite, in the early 1990s (e.g., Mather et al.
1990, 1994; Kogut et al. 1992; Fixsen et al. 1994). These
observations confirming the big bang evolution of the universe are
consistent with the predictions of the standard big bang model
which represents the physically acceptable solutions (Friedmann
1922; Robertson 1935; Walker 1936; Weinberg 1972,1993; Kolb \&
Turner 1994)  of Einstein's equations of the general theory of
relativity. Seemingly, the standard big bang model is superior to
currently popular theories such as superstring theory (Greene
2000) and the inflationary cosmological model (Guth 1997) in
accounting for the observational data associated with the physical
universe. Nevertheless, the occurrence of cosmological singularity
preceding the big bang in the framework of the general theory of
relativity, as shown by the theorems of Penrose (1965), and of
Hawking \& Penrose (1970), raises some of the deepest questions
concerning the initial singularity, a state of infinite density
and temperature.

The problem of cosmological singularity arises because we are
apparently forced to choose only one of the two mutually
complementary fundamental theories of nature. One is Einstein's
general theory of relativity based on the classical concept of
continuous space-time mode of the description of nature, the other
is  quantum theory based on Planck's discovery (1899) of the
universal quantum of action ($h$), which introduces
discontinuities into the law of nature, contrary to the
fundamental principle of classical physics. The general theory of
relativity provides fundamental aspects of our understanding of
natural phenomena in the domain of the large-scale structure of
the universe and  quantum theory in  the domain of  the
microscopic structure of the matter. These two theories, pillars
of modern physics, hitherto have been irreconcilable because of
the basic reasons presented above.  Consequently, to explore the
physics of the big bang and to resolve the problem of the initial
cosmological singularity, the unphysical state of the infinite
density and temperature of the general theory of relativity, we
present a physically acceptable state of finite density and
temperature preceding the big bang.  This is achieved by taking
into account the effects of the uncertainty principle of quantum
theory because of its significant role in the domain of
infinitesimally small region of space-time. Therefore it is
proposed that at the end of the gravitational  collapse of the
universe, without violating the uncertainty principle, there
remained an  immensely large amount of cold condensed energy of
finite density and temperature around the initial singularity
preceding the big bang in a spherically symmetrical state, which
we call the primordial black hole (PBH). In other words, the
primordial black hole is the modified initial singularity
preceding the big bang.

In Sect.2, we present the basic assumptions, and the method of
quantization of the cold condensed energy of the PBH, based on the
principle of dimensional analysis and the reciprocity relation,
which is analogous to that of the unitarity condition of quantum
theory. In Sect.3, we present properties of the PBH in three
parts: 3.1, properties of the  quantized  energy (i.e., the cosmic
particle); 3.2, estimation of the minimum size of the  modified
initial singularity (i.e., PBH); and 3.3,  thermodynamic
properties of the (PBH). In Sect.4, a plausible mechanism of the
big bang, an explosion, analogous to that of a nuclear explosion,
is offered. In Sect.5, we present the results of our calculations
for critical mass density, mean mass density, missing mass as a
function of the Hubble time of expansion (i.e., $\approx$ the age
of the universe) based on our initial conditions. Furthermore we
show that the universe is expanding and closes at the boundary of
the event horizon of the PBH and presumably will recollapse.
Finally in Sect.6, a summary and conclusions are given.
\vspace{3ex}

\noindent \textbf{2. QUANTIZATION OF ENERGY} \vspace{1ex}

\noindent We postulate that during the process of the
gravitational collapse of the universe, in the domain of the
infinitesimally small region of space-time around the initial
singularity, the general theory of relativity breaks down because
such a small region belongs to the realm of quantum theory.
Consequently, without violating the uncertainty principle, there
remained an immensely large and finite density of cold condensed
energy (PBH) in a spherically symmetrical state. According to
Hawking  (1988), ``the higher the mass of the black hole, the
lower the temperature (p.105); a black hole with a mass a few
times that of the sun would have a temperature only one ten
millionth of a degree above absolute zero'' (p.108). Therefore,
even  though we cannot know the internal temperature of the black
hole, we believe that it is reasonable to postulate that the
temperature of the PBH (collapsed state of the universe) with a
mass of $\approx 10^{24}$ times that of the sun would have had a
temperature of $\sim$ absolute zero. Our reasoning for this
postulate is consistent with Penrose's argument based on
thermodynamic considerations that the universe was initially very
regular (i.e., in a state of zero entropy); Chandrasekhar (1990)
likewise agreed with Penrose's view, contrary to those who hold
that the  chaos at the initial singularity was maximal.

The energy of the collapsed state of the universe (PBH) is
quantized in such a manner that each quantum of mass (frozen
energy) is characterized by the fundamental constants of the
general theory of relativity and of quantum theory (namely $G,
\hbar, c$).  For this purpose the principle of dimensional
analysis and the reciprocity relation, analogous to that of the
unitarity condition of  quantum theory,  are incorporated into our
method of quantization so that their physical properties  must be
independent of the units in which they are measured and also must
have lowest value for cosmic unit of mass. Since the dimensions of
$G, \hbar, c$ and mass ($M$) are known, the only simplest
dimensionless combinations which can be formed of them are
$GM^2/\hbar c$ and $\hbar c / GM^2$.  Therefore they must satisfy
the following condition:  \[ \frac{\hbar c}{GM^2} =
\frac{GM^2}{\hbar c}. \]

The only physical solution out of the four solutions $\pm (\hbar
c/G)^{1/2}$ and $\pm i (\hbar c/G)^{1/2}$ for $M$ of the above
relation, we have chosen the real positive value for the mass $M$:
\[ M=\left( \frac {\hbar c}{G} \right)^{1/2} \cong 2.2 \times
10^{-5}~\mathrm{gm}. \]

This is the \emph{lowest} mass of the quantized frozen energy
which we call the cosmic particle. The magnitude of this mass is
identical to that of Planck's mass which he derived  from his
theory of black body radiation (Planck 1899).  Although the values
of their masses are the same, their physical properties are
entirely different. One is an elementary quantum of frozen energy
which has only rest mass while the other is the largest quantum of
radiation energy which has zero rest mass.  The intrinsic spin of
this cosmic particle is considered to be zero because: (i) The
method of quantization of frozen energy is independent of  any
spin; (ii) The method is purely classical and the intrinsic spin
of any  particle is an internal degree of freedom which has no
classical analogue; (iii) If this particle had an intrinsic spin,
it would not have the lowest cosmic unit of mass, rather its
lowest mass would be increased by an amount equivalent to the
rotational energy of its intrinsic spin; and finally (iv) Its
shape must be perfectly spherical because it is in  a state of
absolute zero temperature and lowest energy state. Therefore, we
conclude that most likely its intrinsic spin is zero and it is a
boson. \vspace{3ex}

\noindent \textbf{3. PROPERTIES OF PRIMORDIAL BLACK HOLE}
\vspace{1ex}

\noindent \textbf{3.1 Properties of quantized
particles}\vspace{1ex}

\noindent The minimum size and minimum lifetime of the quantized
cosmic particles, each of mass, $M=(\hbar c/G)^{1/2}$, are
determined within the framework of the uncertainty principle
(Heisenberg 1927). According to this principle, it is impossible
to determine precisely and simultaneously the values of any pair
of canonically conjugate pairs $(q_r,p_s)$ pertaining to a system
and satisfying the usual commutation rules \[ [q_r, p_r] = i\hbar
; \ [q_r,p_s] = 0 \ \mathrm{for} \ r\neq s ; \ [q_r, q_s] = 0 ; \
\mathrm{and} \ [p_r,p_s] = 0.\] In quantitative terms this
principle states that the order of magnitude of the product of
uncertainties in the measurement of two variables must be at least
Planck's constant $h$ divided by $2\pi$ ($\hbar = h/2\pi$).
Therefore, we should have \[\Delta q \cdot \Delta p \geq \hbar ,
\] where $p$ and $q$ are canonically conjugate to each other in
the classical Hamiltonian sense. Profound implications of the
uncertainty principle in physical terms are presented in
complementarity principle (Bohr 1928).  In a four-dimensional
space-time, coordinates of a particle and its corresponding
components of energy-momentum, the uncertainty relations are:  \[
\Delta x \cdot \Delta p_x \geq \hbar ; \ \ \Delta y \cdot \Delta
p_y \geq \hbar ; \ \ \Delta z \cdot \Delta p_z \geq \hbar \ \
\mathrm{and} \ \Delta t \cdot \Delta E \geq \hbar . \] \[
\mathrm{Let} (\Delta x)_{min} = (\Delta y)_{min} = (\Delta
z)_{min} \equiv L_{min} , \]
\[(\Delta p_x)_{max} = (\Delta p_y)_{max} = (\Delta p_z)_{max}
\equiv P_{max} = Mc,\]
\[(\Delta t)_{min} \equiv \tau_m \ \mathrm{and} \ (\Delta E)_{max}
\equiv E = Mc^2, \] therefore:
\[L_{min} \cong \frac {\hbar}{Mc} = \left( \frac {\hbar G}{c^3}
\right) ^{1/2} \approx \ 1.6 \times 10^{-33}~\mathrm{cm~and} \]
\[\tau_m \cong \left(\frac{\hbar}{Mc^2}\right) = \left(
\frac{\hbar G}{c^5}\right) ^{1/2} \approx \ 5.3\times
10^{-44}~\mathrm{sec}. \] $L_{min}(L)$ and $\tau_m$ are
interpreted as the minimum size of localization and minimum
lifetime (time scale) of the quantized cosmic particle. It is
significant that the values of Planck's mass, length, and time
(Planck 1899) are identical to those of the mass, size, and
minimum lifetime of the cosmic particle obtained in the present
investigation. However, we demonstrate later in this article, that
they are entirely different in their properties and  these
different physical properties of the cosmic particle play the most
important role in the big bang theory. Thus our results provide
new significance into Planck's system of units and show that these
units are fundamental not only in the theory of black body
radiation, in the theory of  unification of fundamental
interactions, and in the theory of  superstrings,  but also in the
theory of the big bang origin of the universe. \vspace{3ex}

\noindent \textbf{3.2 Structure of the modified initial
singularity (PBH)}\vspace{1ex}

\noindent The constituents of the modified initial cosmological
singularity (PBH) are cosmic particles, each of mass, $M=(\hbar
c/G)^{1/2}$, and minimum size of localization, $L_{min} = (G\hbar
/c^3)^{1/2}$. These particles are bosons and obey the
Bose-Einstein statistics (Bose 24; Einstein 1924,1925a).
Consequently, without violating the uncertainty principle of
quantum theory and putting all the Bose-Einstein particles in the
lowest energy state as $T \rightarrow 0$, the maximum density of
matter, $\rho_{max}$, in the Primordial black hole (PBH) is given
as a function of $G, \hbar$, and $c$: \[\rho_{max} = \frac
{3}{4\pi} \left(\frac{c^5}{\hbar G^2}\right) \cong 1.2\times
10^{93} ~\mathrm{gm/cm^3}.\]  This result is based on the
assumption that each particle has a hard repulsive core of radius
$\sim 10^{-33}$~cm, consistent with the uncertainty principle,
surrounded by an attractive gravitational force.

The minimum volume and the minimum size $(r_{min})$ of the
structure of the modified initial singularity are determined
taking the total mass of the universe to be about $5.68\times
10^{56}$~gms: This estimation of mass is based on a typical
cosmological model without cosmological constant, compatible with
astronomical observations and with Einstein's conception of
cosmology (Misner, Thorne, \& Wheeler 1973).  We have already
estimated the mass density to be about $10^{93}$~gm/cm$^3$ which
together with the total mass of the universe leads to the minimum
volume $4.76\times 10^{-37}$~cm$^3$ and the minimum size of the
radius ($r_{min}$) $4.8\times 10^{-13}$~cm, about one hundred
thousandth of the size of an atom. These results are significant
because they provide a physical interpretation of the initial
cosmological singularity predicted by the general theory of
relativity. It is not a mathematical point but it has a finite
structure of the order of $10^{-12}$ to $10^{-13}$~cm; and it is
not a 'state' of infinite density but of finite density of the
order of $10^{93}$~gm/cm$^3$. Nevertheless, the distance of the
order of $10^{-13}$~cm is remarkably close to singularity and the
matter density of the order of $10^{93}$~gm/cm$^3$ can be
considered immensely large  and still finite. \vspace{3ex}

\noindent \textbf{3.3 Thermodynamic properties of the PBH}
\vspace{1ex}

\noindent From the available information about the temperatures of
white dwarfs, neutron stars, and black holes (Hawking 1988), and
as well as from Penrose's argument  (Chandrasekhar 1990) we infer,
as stated in Sect.2, that the temperature of the PBH goes to zero
in its final collapsed state. To calculate the entropy  of this
system we utilize Boltzmann's relation between the entropy $S$ and
thermodynamic probability $W$:  \[S = k_\beta \ln W + S_\circ\]
where $k_\beta$ and $S_\circ$ are Boltzmann's and integration
constants respectively.  The first term, $k_\beta \ln W$ is
interpreted as the internal entropy of the matter present inside
the PBH and $S_\circ$ as the external entropy but within the
boundary of the surface area of the event horizon which is
inaccessible to an exterior observer. Since the quantized matter
inside the PBH consists of bosons, its properties can be described
by Bose-Einstein statistics. Therefore the probability $W$ can be
written as (Einstein 1924,1925a): \[ W = \prod_{s} \frac{\left(
N^s +Z^s -1 \right)!}{N^s! \left(Z^s - 1\right)!}.\] Where $N^s$
are the number of particles and $Z^s$ are the number subcells in
the s-th cell. We have pointed out above, all N bosons go into the
lowest energy state as $T\rightarrow 0$, in this limit $N^s = 0$
for all other cells. Therefore $W \rightarrow 1$, $k_\beta \ln W
\rightarrow 0$, and $S \rightarrow S_\circ$. Consequently the
internal entropy of PBH $\rightarrow 0$ as $T \rightarrow 0$. This
result is consistent with the prediction of Einstein in 1925 that
B-E gas satisfies the third law of thermodynamics (Einstein 1925b;
Pais 1982).

The entropy $S_\circ$, inaccessible to an exterior observer, is
calculated within the framework of the theory of black hole
entropy based on the second law of thermodynamics and information
theory developed by Bekenstein  (1973). Therefore \[S = S_\circ =
\frac{1}{2} \left(\frac{\ln 2}{4\pi} \right) k_\beta c^3
\hbar^{-1} G^{-1} A\] where $A$ is the surface area of the event
horizon of the PBH corresponding to the Schwarzschild's radius
$R_{Sch} = 2GM/c^2$ and $A =16\pi M^2 G^2/c^4$. The estimated
value of the entropy of  PBH is found to be  $\approx 1.32\times
10^{107} erg\cdot K^{-1}$ which is extremely large as expected.

Finally the internal pressure in the final phase of the
gravitational collapse (PBH) is evaluated within the framework of
correspondence principle (Bohr 1923) and Newton's gravitational
theory, assuming that the pressure inside the PBH is generated by
an immense, attractive gravitational force alone. At a distance
$r$ from the center of PBH, the pressure $P(r)$ is given by:
\[P(r) = \frac{2\pi}{3}G\rho_{max}^2 \left( r_{min}^2 - r^2
\right); \ \ r \leq r_{min}, \] where $\rho_{max}$ and $r_{min}$
are the maximum matter density and minimum radius of the (PBH)
respectively. The maximum pressure when $r \rightarrow 0$, at the
center, $P(r \rightarrow 0)$ is calculated using the earlier
estimated values of $\rho_{max} \approx 1.2\times
10^{93}$~gm/cm$^3$ and $r_{min} \approx 4.8\times 10^{-13}$~cm.
The result is $P( r\rightarrow 0) \approx 5\times
10^{154}$~dynes/cm$^2$, indeed extremely large. \vspace{3ex}

\noindent \textbf{4. POSSIBLE MECHANISM OF THE BIG BANG}
\vspace{1ex}

\noindent Based on our knowledge of nuclear fission and nuclear
explosives, we give a simplified plausible account of the
mechanism of the big bang. Just as the salient features of the
fission of a nucleus, say uranium by slow neutrons, have been
fully interpreted in terms of the liquid  drop model, originally
by Bohr \& Wheeler (1939); subsequently, this information has been
successfully utilized in the design of various types of explosives
-- from nuclear fission to nuclear fusion hydrogen bombs. The
basic design of a nuclear implosion bomb (Krane 1987), consists of
a solid spherical subcritical mass of the fissionable material
(e.g.,$^{235}$U or $^{239}$Pu) surrounded by a spherical shell of
conventional chemical explosives. An initiator at the center
provides neutrons to start a chain reaction. When the conventional
explosives are detonated in exact synchronization, a spherical
shock wave compresses the fissionable material into a
supercritical state, resulting in an explosion. Now we illustrate
that many of these basic characteristics of a nuclear fission bomb
are naturally present in PBH.

In  the present instance: (1) the PBH which consists of unstable
quantized cosmic particles (bosons) is  analogous to the solid
spherical subcritical mass of the fissionable material;  (2) the
vacuum energy fluctuations caused by the uncertainty principle of
quantum theory surrounding the PBH (within the boundary of its
event horizon) can produce results similar to those of the
conventional explosives surrounding the fissionable material of
the bomb; (3) the unstable cosmic particles, each of  Planck's
mass, are analogous to the nuclei of the fissionable material; and
(4)  the constraint of uncertainty principle and the maximum
gravitational  pressure at the innermost  central  region of  PBH
may trigger the explosion (big bang) analogous to  that  of  the
initiator at the center of the bomb which provides neutrons to
begin the chain reaction resulting in explosion. In  an explosion
of a nuclear fission bomb, only a very small fraction of mass of
each nucleus of fissionable material is transformed into kinetic
energy while in the explosion of the PBH, total mass ($M \cong 2.2
\times 10^{-5}$~gms) of each unstable cosmic particle is converted
into kinetic energy resulting in temperature of $10^{32}$~K, the
currently accepted temperature at the beginning of the big bang.
The total mass of the PBH consisting of nearly $10^{61}$ cosmic
particles, each of mean lifetime (decay rate) of the order of
$\tau_m \approx 5.3\times 10^{-44}$~sec, is transformed into
kinetic energy $\approx 10^{77}$~erg, practically in less than
$10^{-40}$~second resulting in the biggest explosion, the big
bang. In the following section, we examine the aftermath of the
big bang. \vspace{3ex}

\noindent \textbf{5. THE COSMIC MEAN MASS DENSITY AND THE HUBBLE
TIME OF EXPANSION} \vspace{1ex}

\noindent Our discussions of the cosmic mean mass density and the
Hubble time of expansion are based on the assumptions that the
universe is homogeneous and isotropic. Further, Einstein's
cosmological principle requires that the world view of any
observer relative to himself must have spherical symmetry about
himself. Therefore let us consider a sphere of galaxies of radius
$R(t)$ with the observer at the origin. According to a theorem of
Newton which remains valid also in the framework of general theory
of relativity as shown by Birkhoff (1927); the total energy $E$ of
a particle of mass $m$ on the surface of the  sphere of radius
$R(t)$ with a uniform mass density $\rho(t)$ is (Weinberg 1972,
p.475):
\begin{eqnarray}
    E & = & \frac{1}{2}m \frac{\left|X(t_\circ)\right|^2}{R^2(t_\circ)}\left[
            \left( \frac{dR(t)}{dt} \right)^2 - \frac{8\pi G}{3} \rho(t) R^2(t)
            \right] \nonumber \\
      & = & \frac{1}{2}m \frac{\left|X(t_\circ)\right|^2}{R^2(t_\circ)}R^2 H^2
            \left[1-\Omega \right],
\end{eqnarray}
where $H=\dot{R}/R$ (where $\dot{R} \equiv dR/dt$), is the Hubble
constant which determines the expansion rate of the universe; the
total energy, $E$, remains constant as the universe expands: and
$\Omega$ is the ratio of the density $\rho$ to the critical
density $\rho_c$:  \be \Omega \equiv \frac{\rho}{\rho_c}; \ \
\mathrm{and} \ \ \rho_c \equiv \frac{3H^2}{8\pi G}. \ee

The equation (1) is equivalent to the following Friedmann equation
of the cosmology, a solution of Einstein's equations of the
general theory of relativity without the cosmological constant,
based on the Robertson-Walker metric (Robertson 1935; Walker 1936;
Weinberg 1972; Kolb \& Turner 1994)  \[k+\dot{R}^2 = \frac{8\pi
G\rho R^2}{3}, \]  or in terms of Hubble Constant $H$ and
$\Omega(\rho/\rho_c)$:  \be k = -R^2H^2 [1-\Omega] \ee Provided
the constant energy $E$ and the constant $k$ are given by the
relation:  \be E = -\frac{1}{2}m\frac{\left|X(t_\circ)
\right|^2}{R^2(t_\circ)} k. \ee  Equation (1), which is equivalent
to the Friedmann Eq.(3) of  cosmology, is valid for all times.
However note that $\rho$, $\rho_c$, $H$, and $\Omega(\rho/\rho_c)$
are not constant but change as the universe expands. We shall now
consider three cases of Eq.(1) for the future evolution of our
universe: (i) For $k=-1$, Eq.(4) shows that the total energy $E$
is positive-definite, and Eq.(1) shows that the expansion velocity
$\dot{R}$ never can go to zero, $\Omega$ must be less than one,
consequently if the universe is presently expanding it must
continue to expand forever (i.e., the universe is open); (ii) For
$k=+1$, Eq.(4) shows that the total energy $E$ is
negative-definite, and therefore Eq.(1) shows that $\Omega$ must
always remains greater than one, then at a finite time after the
big bang origin of the universe, the universe will achieve a
maximum expansion radius $R$, the expansion velocity $\dot{R}$
goes to zero, and then it will begin to recollapse (i.e., the
universe is closed); and finally (iii) For $k = 0$, Eq.(4) shows
that the total energy, $E$, of the cosmic particle is zero, and
Eq.(1) for this case shows that $\Omega = 1$, and if the universe
is presently expanding it must continue to expand forever since
expansion velocity, $\dot{R}$, asymptotically approaches zero as
$\rho(t)R^2(t) \rightarrow 0$ as $R(t)\rightarrow \infty$;
assuming $\rho(t)R^3(t)$ is constant which expresses the
conservation of mass in the Newtonian considerations. From these
results it is evident that the hot big bang model, more properly
the Friedmann-Robertson-Walker cosmology (standard model of
cosmology) does not give a definite prediction as to how the
universe will end, i.e., whether the universe is open or closed.
Furthermore, it does not attempt to describe what might have
existed preceding the big bang and how it might have ``banged''.
In essence, the standard cosmological model, although the most
successful model of cosmology, describes the evolution of the
universe, the aftermath of the big bang, from unknown initial
conditions and ends with the unknown final future of the universe.

Therefore it is of major interest to determine the critical mass
density, $\rho_c$, the present mean mass density, $\rho_p$, and
the ratio of the mean density to the critical density, $\Omega_p =
\rho_p/\rho_c$, as a function of the Hubble time of expansion
based on our postulate that all the matter in the universe, namely
$M_p \cong 5.68\times 10^{56}$~gm (Misner et al. 1973) existed
preceding the big bang. We show in the Appendix that the total
mass which includes all forms of energy is conserved in the
standard model of cosmology. The currently popular accepted value
for the Hubble constant, $H$, based on 15 Kilometers per second
per million light year is: \be H \cong (6.31 \times 10^{17} \
\mathrm{sec})^{-1}. \ee  Case I.  The critical mass density,
$\rho_c$, corresponding to the above value of $H$; the Hubble time
of expansion, $T_H = H^{-1} \approx 20$~billion years; and the
Hubble radius of expansion $R=c/H \approx 1.89\times 10^{28}$~cm
is given:  \be \rho_c = \frac{3H^2}{8\pi G} \approx 4.5\times
10^{-30} \ \mathrm{gm/cm^3}. \ee  Case II.  Consistent with our
postulate that all matter present in the universe, namely $M_p =
5.68\times 10^{56}$~gm, existed preceding the big bang in the
primordial black hole, it must remain the same after the big bang
as $R(t)$ varies with time, as required by the law of conservation
of mass (see Appendix).  Consequently the mean mass density,
$\rho_p$ in the present work, corresponding to the Hubble time of
expansion $T_H = H^{-1} \approx 20$~billion years and the initial
mass, $M_p= 5.68\times 10^{56}$~gm is: \be \rho_p =
\frac{3M_p}{4\pi c^3 H^{-3}} \approx 2.01\times 10^{-29} \
\mathrm{gm/cm^3} \ee compared to the critical mass density,
$\rho_c$, for $T_H \approx$~20~billion years is $\approx 4.5\times
10^{-30}$~gm/cm$^3$.  Therefore the universe must be closed, since
$\Omega_p = \rho_p/\rho_c > 1$ and it must also be expanding.

Now let us determine when the expansion of the universe stops. For
this purpose we are going to prove that when $E=0$, in Eq.(1) or
$k=0$, in Eq.(3), and $\Omega = \Omega_p = 1$ (i.e., when $\rho_p
= \rho_c$), the expansion of the universe stops and closes.

Let $H_{min}$ be the minimum Hubble constant; $R_{max}$, the
maximum radius of expansion; $\rho_{p,min}$, the minimum mass
density; and $\rho_{c,min}$, the minimum mean critical mass
density, then since $\rho_{p,min} = \rho_{c,min}$, we obtain:
\[\frac{3M_p}{4\pi R^3_{max}} = \frac{3H^2_{min}}{8\pi G}; \]
or, equivalently \[\frac{3M_p}{4\pi R^3_{max}} = \frac{3c^2}{8\pi
GR^2_{max}}; \ \mathrm{since} \ H_{min}=\frac{c}{R_{max}}.\]  The
above relation gives:  \be R_{max} = \frac{2GM_p}{c^2} = R_{Sch},
\ee where $R_{Sch}$ is the Schwarzschild radius of the primordial
black hole (PBH).  Consequently, the expansion of the universe
must close at the boundary of the event horizon, $R(t)\leq
R_{Sch}$, of the PBH after the big bang since nothing can escape
from it. This result is consistent with a mathematical proof given
by Pathria (1972). Therefore  the maximum Hubble time,
$T_{H,max}$, of expansion of the universe aftermath of the big
bang  in the present case is:  \be T_{H,max} = H_{min}^{-1} \leq
\frac{R_{Sch}}{c} = \frac{2GM_p}{c^3} \approx 89 \ \mathrm{billion
\ years}.\ee

Now we discuss the question of mass difference (i.e., the missing
mass) as a function of the Hubble time of expansion $t$, between
the total critical mass, $M_c(t)$, and the total initial mass,
$M_p$, which must remain constant, as $R(t)$ varies with time
consistent with the law of conservation of mass. Since the Hubble
time of expansion, $T_H = H^{-1}(t) =R(t)/c$, the total critical
mass, $M_c(t)$, as a function of time $t$, is given by \be M_c(t)
= \frac{4}{3}\pi R^3(t) \cdot \frac{3H^2(t)}{8\pi G} =
\frac{c^3}{2GH(t)}. \ee  Therefore the percentage of missing mass
as a function of expansion time $t$, aftermath of the big bang is
given by: \be \frac{[M_p - M_c(t)]}{M_p}10^2. \ee

We now proceed to calculate the critical mass density $\rho_c(t)$,
the present mean mass density $\rho_p(t)$,  the critical total
mass $M_c(t)$, and the percentage of missing mass as a function of
the Hubble time of expansion, $T_H = H^{-1}(t)$ by using the
Equations (6), (7), (10), and (11) respectively; and also
$\Omega_p$, the ratio of the present mean mass density $\rho_p$ to
the critical mass density $\rho_c$. The results of our
calculations are given in Table 1. It is clear from the Table 1
that as the universe expands, with increasing value of the Hubble
time of expansion, $T_H$:  $\rho_c(t)$, $\rho_p(t)$,
$\Omega_p(t)$, and the percentage of missing mass, all of them
decrease. However, the total critical mass $M_c(t)$ increases and
the initial mass $M_p$ remains the same consistent with the law of
conservation of mass-energy.  By the time the expansion closes at
the boundary, $R(t) \leq R_{Sch}$, of the event horizon of the
primordial black hole: $\rho \approx 2.27\times
10^{-31}$~gm/cm$^3$; $M_c(t) \rightarrow M_p \approx 5.68\times
10^{56}$~gm; $R(t) \rightarrow \leq R_{Sch} \approx 8.4\times
10^{28}$~cm; $T_{H,max} \rightarrow \approx 89\times 10^9$~years;
and the missing mass is zero. It is also seen from the Table 1
that corresponding to the currently popular accepted value for the
Hubble constant, $H \approx (6.31\times 10^{17} \
\mathrm{sec})^{-1}$ (i.e., the age of the universe, $H^{-1} = T_H
\approx  20\times 10^9$~years): $\rho_c = 4.5\times
10^{-30}$~gm/cm$^3$; $\rho_p = 2.01\times 10^{-29}$~gm/cm$^3$;
$\Omega_p = 4.47$; the total critical mass $Mc = 1.278\times
10^{56}$~gm compared to the total initial mass $M_p = 5.68\times
10^{56}$~gm (i.e., the missing mass $\approx 77.70$ percentage);
and the universe is expanding. However, note that the universe
remains inside the event horizon of the primordial black hole
(i.e., the modified initial singularity).
\\
\\

\begin{flushleft}
TABLE 1 \vspace{0.5cm}

Initial mass of the PBH, $M_p = 5.68\times 10^{56}$~gm (observed
mass)

$T_H(H^{-1}) = T_\circ \times 10^9$~years (Hubble time of
expansion)

$\rho_c = \rho_\circ \times 10^{-28}$~gm/cm$^3$ (critical density)

$M_c = M_\circ \times 10^{55}$~gm (corresponds to critical
density)

$\rho_p = \rho \times 10^{-28}$~gm/cm$^3$ (density in the present
work)

$M_p = M \times 10^{56}$~gm (initial mass preceding the big bang)

$\Omega_p = \rho_p/\rho_c$ [in the present work $\Omega_p$ is a
function of $T_H(H^{-1})$]
\end{flushleft}

\[ \begin{array}{ccccccc}
    \ \ \ \ T_\circ \ \ \ \ & \ \ \ \ \rho_\circ \ \ \ \ & \ \ \ \ M_\circ \ \ \ \ & \ \ \ \ \rho \ \ \ \ & \ \ \ \ M \ \ \ \ & \ \ \ \ \Omega_p \ \ \ \ &
    \begin{array}{c}\mathrm{Percentage} \\ \mathrm{of \ missing} \\ \mathrm{mass} \end{array}
    \\
    3   & 2.006   &  2.01   & 59.185  &  5.68  &  29.50  &  96.50\\
    5   & 0.719   &  3.19   & 12.78   &  5.68  &  17.77  &  94.40\\
    10  & 0.1798  &  6.387  & 1.598   &  5.68  &  8.89   &  88.75\\
    15  & 0.0799  &  9.581  & 0.473   &  5.68  &  5.92   &  83.13\\
    20  & 0.045   &  12.777 & 0.201   &  5.68  &  4.47   &  77.70\\
    30  & 0.01997 &  19.16  & 0.0591  &  5.68  &  2.96   &  66.30\\
    40  & 0.01124 &  25.6   & 0.02497 &  5.68  &  2.22   &  54.93\\
    50  & 0.00719 &  31.93  & 0.01278 &  5.68  &  1.78   &  43.80\\
    60  & 0.00499 &  38.32  & 0.0074  &  5.68  &  1.48   &  32.50\\
    80  & 0.00289 &  51.1   & 0.0031  &  5.68  &  1.08   &  10.03\\
    89* & 0.00227 &  56.8   & 0.00227 &  5.68  &  1.00   &  0.00
   \end{array} \]

\noindent *89 Billion years corresponds to the Hubble time of
expansion for the length of Schwarzschild radius ($R_{Sch}/c$)
event horizon of the PBH.

\noindent Note:  Initial mass-energy remains constant before and
after the big bang.  Expansion closed at the boundary of the event
horizon. The mass density is interpreted as the total energy
divided by $c^2$.\vspace{3ex}

\noindent \textbf{SUMMARY AND CONCLUSIONS} \vspace{1ex}

\noindent The big bang theory poses some of the deepest questions
in physics today, for example: (1) what, if anything, preceded the
big bang, and how did it happen?; and (2) how  will the universe
end, i.e., is the universe  open, flat, or closed? To search for
the most plausible physical solution, we have postulated that
during the process of gravitational collapse at the
infinitesimally small region of space-time, the general theory of
relativity breaks down because such a small region belongs to the
realm of quantum theory. Consequently, at the end of the
gravitational collapse, instead of the initial cosmological
singularity with infinite density and temperature, we now have an
extended infinitesimally small structure of cold condensed energy
(PBH) with finite density and temperature, preceding the big bang.
This result has been achieved by taking into account the effect
the uncertainty principle of quantum theory.

The frozen energy of the collapsed state has been quantized in
such a manner that each quantum of energy (mass) called cosmic
particle, is characterized by the fundamental constants, $G$, $c$,
$\hbar$, of the general theory of relativity and quantum theory.
The present method of quantization predicts four solutions, $\pm
(\hbar c/G)^{1/2}$  and  $\pm i(\hbar c/G)^{1/2}$ for the mass of
the cosmic particle with intrinsic spin zero.  In the present work
we have chosen the real positive value for the mass of the cosmic
particle, $M = (\hbar c/G)^{1/2} \cong 2.2\times 10^{-5}$~gms. We
have also determined its minimum size of localization, $L_{min}
\cong 1.6\times 10^{-33}$~cm, and minimum lifetime $t_m \cong
5.3\times 10^{-44}$~sec within the framework of Heisenberg's
uncertainty principle. It is significant that the values of
Planck's system of units:  mass, length, and time, derived by
Planck (1899), when neither theory of relativity nor quantum
theory had been discovered, are identical to those obtained in the
present  work within the framework of these two fundamental
theories of the twentieth century. However, they have different
physical properties: In Planck's work, mass $M_p \equiv h
\nu_{max}/c^2$ is kinetic energy with rest the mass zero and,
$\nu_{max}$ represents the largest frequency of his classic theory
of black body radiation; length $L_p \equiv \lambda_{min}$,
$\lambda_{min}$ represents the minimum wavelength of the
radiation; and time $T_p \equiv 1/\nu_{max}$, the period of
vibration of wavelength $\lambda_{min}$ (i.e., the minimum time
scale). In the present investigation, $M$ represents the rest mass
of the quantized cosmic particle; length $L$($L_{min}$) represents
the minimum size of localization of the cosmic particle; and time
$T$($\tau_m$) represents the minimum lifetime (i.e., the minimum
time scale) of the cosmic particle.

The above properties of the cosmic particles (bosons),
constituents of the extended structure of the cosmological
singularity (PBH); have been utilized to determine the physical
properties of the modified initial singularity (PBH), namely: (1)
the maximum matter density, $\rho_{max} \cong 1.2\times
10^{93}$~gms/cm$^3$; (2) the minimum volume $\cong 4.76\times
10^{-37}$~cm$^3$; (3) the minimum radius, $r_{min} \cong 4.8\times
10^{-13}$~cm; (4) the entropy, $S \cong 1.37\times
10^{107}$~erg$\cdot$K$^{-1}$; and (5) the maximum gravitational
pressure at the center, $P(r\rightarrow 0) \approx 5\times
10^{154}$~dynes/cm$^2$, of the PBH in its final phase of the
gravitational collapse state. Based on the properties of the
cosmic particles and those of the PBH, we have presented in Sect.4
a  most reasonable account of the mechanism of the explosion of
the matter of the PBH, resulting in a temperature of $10^{32}$~K
at the beginning of the big bang and energy evolved $\approx
10^{77}$~erg.

Furthermore, we have calculated the cosmic mean mass density
$\rho_p(t)$, the critical mass density $\rho_c(t)$, the ratio of
the mean mass density to the critical mass density $\Omega_p$, and
the missing mass as a function of the Hubble time of expansion
$T_H = H^{-1}(t)$, based on our postulate that all the matter
present in the universe, namely $M_p = 5.68\times 10^{56}$~gm
(Misner et al. 1973), existed preceding the big bang inside the
primordial black hole (PBH).  The results of our calculations are
presented in Table 1.  It is clear from the Table 1, that as the
universe expands, with increasing value of the Hubble time $T_H =
H^{-1}(t)$: $\rho_c(t)$, $\rho_p(t)$, $\Omega_p(t)$, and the
missing mass decrease; whereas the total critical mass $M_c(t)$
increases and the initial mass $M_p$ remains the same as required
by the law of conservation of mass (energy).  It is relevant to
point out here that the percentage of missing mass given in Table
1, is the percentage of mass difference between the initial mass
$M_p$ and the total critical mass $M_c(t)$ as explained in Sect.5,
Eq.(11). Consequently, if the total observed visible mass density
$\rho_{vd}(t) < \rho_c(t)$, then the percentage of missing mass
given in Table 1 will be further increased.  It should be also
noted that $\rho_c(t)$ is independent of the initial mass $M_p$
(i.e., the initial condition) and depends only on the Hubble time
of expansion whereas $\rho_p(t)$ depends  both on the Hubble time
of expansion and the initial mass $M_p$.

Furthermore, it is seen from Table 1 that corresponding to the
currently popular accepted value for the Hubble constant, $H \cong
(6.31\times 10^{-17} \ \mathrm{sec})^{-1}$ (i.e., the age of the
universe $H^{-1} = T_H \approx 20\times 10^9$~years): $\rho_c =
4.5\times 10^{-30}$~gm/cm$^3$; $\rho_p = 2.01\times
10^{-29}$~gm/cm$^3$; $\Omega_p = 4.56$; the total critical mass
$M_c = 1.278\times 10^{56}$~gm compared to the initial mass $M_p =
5.68\times 10^{56}$~gm (i.e., the missing mass $\approx 77.70$
percentage); and the universe is expanding.  Finally, we have
shown in Sect.5, that the universe closes at $R(t) \rightarrow
\leq R_{Sch}$, Schwarzschild  radius of the event horizon of the
primordial black hole and presumably will recollapse (i.e., the
universe always remains inside the event horizon of the PBH).
Unquestionably, our results and inferences are significantly
different from those of the predictions of  the
Friedmann-Robertson-Walker model of cosmology (also known as
standard model of cosmology) (Friedmann 1922; Robertson 1935;
Walker 1936; Weinberg 1972; Kolb \& Turner 1994). It is not
surprising because the standard cosmological model, although the
most successful model of modern cosmology, describes the evolution
of the universe, aftermath of the big bang, from unknown initial
conditions and ends with the unknown final future of the universe
as expected from the laws of classical and quantum theories.
\\
\\

\noindent \textbf{ACKNOWLEDGEMENTS} \vspace{1ex}

\noindent The author wishes to thank Prof. Malvin A. Ruderman and
Prof. Englebert L. Schucking for their helpful discussions in the
early stage of this work and in particular to Dr. William J.
Marciano for his valuable comments and discussions concerning this
manuscript.
\\
\\

\noindent \textbf{APPENDIX.  CONSERVATION OF MASS (ENERGY) IN
STANDARD MODEL} \vspace{1ex}

\noindent Friedmann's equation of energy conservation in
Robertson-Walker metric of Einstein's equations of general
relativity (Weinberg 1972), 472, equation (15.1.21) is, \[
\frac{dp}{dt} R^3 = \frac{d}{dt} \left\{R^3(\rho + p)\right\}; \]
or, equivalently
\[ d\left(\frac{4}{3}\pi R^3 \rho \right) = -p d\left(\frac{4}{3}
\pi R^3 \right); \] or,

\begin{flushright}$dM = -p dV.$ \hspace{5.7cm} (A1)
\end{flushright}

Since in Friedmann's equation of energy conservation, $c=1$,
$U\equiv E = Mc^2$, and according to the Equipartition theorem, $U
= E = k_\beta T$, where $k_\beta$ is the Boltzmann's constant and
$T$ is the temperature, therefore equation (A1) can also be
written as,

\begin{flushright} $dM + \frac{p}{c^2}dV = 0$ \hspace{5.5cm} (A2)

$dU + p dV = 0$ \hspace{5.7cm} (A3)

$k_\beta dT + p dV = 0.$ \hspace{5.4cm} (A4) \end{flushright}

Note that equation (A3) is the familiar form of the first law of
thermodynamics and it represents the principle of conservation of
energy for an isolated system: $dU$ represents the change in
internal energy of the system and $pdV$ represents the work done
by the system.  They are equal in magnitude and opposite in sign.

The above equations demonstrate that during the process of the
evolution of the universe, decrease of mass or energy or
temperature is compensated by the increase in volume due to
expansion of the universe and thus energy is conserved.
\\
\\

\noindent \textbf{REFERENCES}
\begin{description}
    \item Alpher R. A., Bethe H., Gamow G., 1948, Phys. Rev., 73, 803
    \item Alpher R. A., Follin J. W. Jr., Herman R. C., 1953, Phys. Rev.,
        92, 1347
    \item Bekenstein J. D., 1973, Phys. Rev. D, 7, 2333
    \item Birkhoff G. D., 1927, Relativity and Modern Physics, 2nd ed.. Harvard Uni. Press,
        Cambridge, Massachusetts, p.253
    \item Boesgaard A. M., Steigman G., 1985, ARA \& A, 23, 319
    \item Bohr N., 1923, Z. Phys., 13,117
    \item Bohr, N., 1928, Nature, 121, 580
    \item Bohr  N., Wheeler J. A., 1939, Phys. Rev., 56, 426
    \item Bose S. N., 1924, Z. Phys., 26, 178
    \item Chandrasekhar S., 1990, Selected Papers, Vol. 5, Relativistic
        Astrophysics. Univ. Chicago Press, Chicago, p. 577
    \item Einstein A., 1924, Sitzungsber. Phys.-Math. Kl. p.261
    \item Einstein A., 1925a, Sitzungsber. Phys.-Math. Kl. p.3
    \item Einstein A., 1925b, Sitzungsber. Phys.-Math. Kl. p.18
    \item Fixsen D. J. et. al., 1994, ApJ, 420, 445
    \item Friedmann A., 1922,  Z. Phys.,10, 377; 1924, Z. Phys., 21, 326
    \item Gamow G., 1946, Phys. Rev., 70, 572
    \item Gamow G., 1948, Phys. Rev., 74, 505
    \item Greene B., 2000, The Elegant Universe. Vintage Books, A division
        of Random House, Inc. New York; see for a detailed discussion of
        the superstring theory.
    \item Guth A. H., 1997, The Inflationary Universe. Addison-Wesley,
        Reading, Massachusetts; see for a detailed discussion of the
        inflationary cosmological model
    \item Hawking S. W., 1988, A Brief History of Time. Bantam Books, New
        York
    \item Hawking  S. W., Penrose R., 1970, Proc. Roy. Soc. Lond., A314, 529
    \item Heisenberg W., 1927, Z. Phys., 43, 172
    \item Hubble E. P., 1929, Proc. Nat. Acad. Sci. U.S., 15, 169
    \item Kogut A. et al., 1992, ApJ, 401, 1
    \item Kolb E. W., Turner M. S., 1994, The Early Universe, Paperback Edition.
        Addison- Wesley, Reading, Massachusetts
    \item Krane K. S., 1987, Introductory Nuclear Physics. John Wiley \&
        Sons, Inc., New York, p. 521
    \item Mather J. C. et al., 1990, ApJ, 354, L37
    \item Mather J. C. et al., 1994, ApJ, 420, 439
    \item Misner C. W., Thorne K. S., Wheeler J. A., 1973, Gravitation. W.H.
        Freeman \& Co., San Francisco, CA, p. 738
    \item Pais A., 1982, Subtle is the Lord: the Science and the Life of
        Albert Einstein. Oxford University Press, New York,  p. 431
    \item Pathria R. K., 1972, Nature, 240,298
    \item Penrose R., 1965, Phys. Rev. Lett., 14, 57
    \item Penzias A. A., Wilson R. W. A., 1965, ApJ, 142, 419
    \item Planck M., 1899, Sitzungsberichte, Deut.Akad.Wiss.Berlin, Kl.
        Math.-Phys. Tech., pp440-480
    \item Robertson H. P., 1935, ApJ, 82, 248; 1936, ApJ, 83, 187, \& 257
    \item Walker A. G., 1936, Proc. Lond. Math. Soc., 42, 90
    \item Weinberg  S., 1972, Gravitation and Cosmology. John Wiley \& Sons,
        New York; 1993, The First Three Minutes: A Modern View of the
        Origin of the Universe, 2d  ed..  Basic Books, A Division of
        Harper Collins, New York
\end{description}

\end{document}